\documentclass[smallextended]{svjour3} 
\usepackage{graphicx}
\usepackage{color}
\usepackage[utf8]{inputenc} 
\usepackage[T1]{fontenc}    
\usepackage{hyperref}       
\usepackage{url}            
\usepackage{booktabs}       
\usepackage{amsfonts}       
\usepackage{nicefrac}       
\usepackage{microtype}      
\usepackage{color}

\title{How Downwards Causation Occurs in Digital Computers}

\author{
George Ellis 
   \and
   Barbara Drossel
  }

\institute{G  Ellis \at
              Department of Mathematics\\
  University of Cape Town\\
  Rondebosch, Cape Town 7701\\
 \email{george.ellis@uct.ac.za}           
           \and
B Drossel \at Institute of Condensed Matter Physics\\
  Technische Universit\"at Darmstadt\\
  Hochschulstr. 6, 64289 Darmstadt \\
  \email{drossel@fkp.tu-darmstadt.de}           
}

\date{Received: date / Accepted: date}
\begin{document}
\maketitle


\begin{abstract}
Digital computers carry out algorithms coded in high level programs. These abstract entities  determine what happens at the physical level: they control whether electrons flow through specific transistors at specific times or not, entailing downward causation in both the logical and implementation hierarchies. This paper explores how this is possible in the light of the alleged causal completeness of physics at the bottom level, and highlights the mechanism that enables strong emergence (the manifest causal effectiveness of  application programs) to occur. Although synchronic emergence of higher levels from lower levels is manifestly true, diachronic emergence is generically not  the case; indeed we give specific examples where it cannot occur because of the causal effectiveness of higher level variables.
\end{abstract}

\section{Introduction}\label{sec:intro}
At a foundational level, physics is based in time reversible Hamiltonian dynamics. 
Emergence of complex systems involves symmetry breaking, as famously pointed out by Anderson \cite{Anderson}, which occurs for example when contextual effects introduce time dependent constraints in the Hamiltonian. We will  discuss this issue here in the context of the functioning of digital computers, because this provides a clear case where we can in principle fully understand what is going on at every level of the emergent hierarchy \cite{Tannen} - after all, we have designed and built them. This is a key issue in terms of the relation between fundamental physics and emergent entities \cite{Anderson}; we agree with the illuminating discussion by Leggett \cite{Leggett1992} in this journal (see Section \ref{sec:disconnect} below).\\

The issue of whether strong emergence is possible is a highly contested topic, see for example \cite{Hohwyand},  \cite{Humph}, and \cite{GibbHendryLancaster} for discussion. 
This is closely linked to the issue of downwards causation (\cite{Ellisetal2012}, \cite{Ellis2016}), because if higher emergent levels have real causal powers in their own right, as claimed by \cite{Noble_2012} in the case of biology, they must have the ability to act down to lower levels in order to shape lower level dynamics so as to meet higher level needs.
The purpose of the present paper is to show how this occurs in digital computers, where algorithms embodied in computer programs, together with data,  are causally effective  by shaping  outcomes at the electron level.  
This is what enables the causal power of computer programs in the physical world, as for instance in car manufacture, landing  an aircraft, playing music,  playing
chess, or searching the web \cite{McCormack_algorithms}. Thus strong emergence certainly takes place in those cases: these outcomes cannot be derived in a solely bottom-up way, because they enable conditional branching logic to have causal power, for example (``IF \{\textit{aircraft too high}\} THEN \{\textit{reduce engine power}\}''), and so on, where ``too high'' is a highly contextual variable depending on wind, aircraft type, load, and so on.\\

Here `causation' is understood  in a counter-factual way \cite{Menzies}: if the application program were different, which would be the case if a different algorithm were employed (for example computing airflow over wings, rather than the next move  in a chess game), there would be different flows of electrons through gates at the transistor level. Thus both the dynamics of the gates (the time-dependent way they implement Boolean logic) and of the electrons (which carriers flow through which semi-conductor materials at what time) are  controlled jointly by the application programs implemented in the computer, the data loaded, and the time the program is turned on. These are in turn determined by the computer operator in the social setting which is the context in which this all happens, which social context also led to the existence of the computer in the first place. Physics is not causally complete by itself; it is only causally complete in this full context (see \S\ref{sec:conclude}).\\

It has been claimed in Chapter 2 of \cite{Ellis2016} that downwards causation takes place in digital computers, and this is implicit in H A Simon's discussion of technological systems \cite{SimonHA},  Tannebaum's description of hierarchical computer organisation \cite{Tannen}, and Mellisinos' presentation of how the logical and physical levels in digital computers interrelate \cite{Mellis}. However none of these writings comment on how the branching logic of algorithms can be realised on the basis of the  alleged underlying Hamiltonian physics, whose unitary nature is the essential basis for claims that supervenience - the statement that a specific lower level state uniquely  determines a specific higher level state, and any difference in higher level state must be reflected in a difference in lower level states -  undermines the possibility of strong emergence.\\

In this paper, \S\ref{sec:modular_Hierach} sets the stage by discussing the relevant hierarchical structures. Sections \S\ref{sec:Branching_LOgica}-\S\ref{sec:variables_nature} identify the mechanisms whereby branching of logical variables can reach down to cause branching of physical dynamics at the electron level, with \S\ref{sec:Branching_LOgica} considering branching logic at all levels,  \S\ref{sec:contextual_branch_Transistor} looking at contextual branching at the transistor level, and \S\ref{sec:variables_nature} the nature of variables. While synchronic supervenience (that is, supervenience occuring at one time)  can be claimed to always occur in digital computers, in many cases diachronic supervenience (that is, supervenience occurring at an earlier time in relation to outcomes at a later time) cannot occur in many cases, as discussed in \S\ref{sec:supervene}. \\

These considerations show how  abstract structures can have causal powers (\S\ref{sec:causal_power}) and 
physics at the lowest level is not causally complete (\S\ref{sec;Causal_completeness}). This is analogous to the way such downward causation  happens in biology, as discussed in \cite{Ellis_Koppel}. However the mechanism whereby the underlying microscopic processes at the electron level get conscripted to fulfill higher level purposes is different in the two cases (\S\ref{sec:difference_biology}). The outcome is that  all levels are equally real (\S\ref{sec:levels_all_real}), and personal and social needs determine what actually happens in terms of electron flows in transistors in digital computers (\S\ref{sec:social_connection}).
\section{Modular hierarchical structures}\label{sec:modular_Hierach}
Like all truly complex systems, digital computers are \textit{Modular Hierarchical Structures} in both physical and logical terms. Because they are structured, higher levels can constrain and control lower level dynamics \cite{constraints}. In order to achieve complex emergent outcomes, that structure must be both hierarchical (\S\ref{sec:orthog_hierarch}) and modular (\S\ref{sec:modularity}). A key feature of such structures is multiple realisability of higher level structure and functions at lower levels (\S\ref{sec:Multiple_Realis}).

\subsection{The orthogonal hierarchies} \label{sec:orthog_hierarch}
In digital computers, there is a logical hierarchy and an implementation hierarchy (\cite{Ellis2016}:\S2). The latter implements the former. Upwards emergence of higher levels and downwards constraints on lower levels take place in both cases.

\paragraph{The logical hierarchy (software)} This tower of `virtual machines' $M_I$, each with associated language $L_I$, is discussed in depth in  (\cite{Tannen}:pp.2-8), and is represented in Table 1.
\newpage
\vspace{0.1in}

\begin{tabular}{|c|l|c|c|}
	\hline \hline
Level	&  Logic & Downward Process & Agent \\ 
	\hline \hline
$M_6$	& $L_6$: Application program  &  $\Downarrow$ Logic and data & Program code\\  
\hline
$M_5$	& $L_5$:  High level language  &  $\Downarrow$ Translation: Compiler & Syntax, Semantics\\  
	\hline 
$M_4$	& $L_4$:  Assembly language  & $\Downarrow$ Translation: Assembler & Syntax, Semantics \\ 
	\hline 
$M_3$	& $L_3$: Operating system  level & $\Downarrow$ Partial Interpretation & Operating system \\ 
	\hline 
$M_2$	&  $L_2$: Machine language %
& $\Downarrow$ Interpretation & ISA	 \\ 
\hline 
$M_1$	& $L_1$: Microprograms   & $\Downarrow$ Directly executed & Instructions \\ 
	\hline 
$M_0$	&  Digital logic level (Binary) &  Hardware &Gates  \\ 
	\hline \hline
\end{tabular} 

\vspace{0.1in}
\noindent	\textbf{Table 1}:\label{Table1} \textit{Multilevel machine with a tower of virtual machines $M_I$ each with a language $L_I$. The lowest logical level is $M_1$ with machine language $L_1$, with binary instructions directly executed by electronic circuits. ISA is the Instruction Set Architecture.} Adapted from \cite{Tannen}, Figs. 1-1, 1-2.\footnote{Tanenbaum  \cite{Tannen} does not distinguish between the logical and implementation hierarchies as we do here. We view this as a key distinction.}

\vspace{0.1in}
\noindent Each language is characterised by its Syntax and its Semantics. The user interacts with the application program level $M_6$, which might be a word processor, internet browser, image processor, etc; 
or with the high level language level $M_5$: languages such as FORTRAN, C, JAVA, PYTHON, etc. Compilers or assemblers chain the high level logic down to the microprogramming level; the resulting binary code can be directly executed by the hardware \cite{Tannen}. 

\paragraph{The implementation hierarchy (hardware)} 
\label{sec:hierarch_impl}
This hierarchy, see Table 2, is implied by \cite{Mellis} and \cite{Tannen}
. It is what enables the logical hierarchy to influence physical reality. However there is not a 1-1 correspondence between the levels in the two hierarchies.

\vspace{0.1in}

\begin{tabular}{|c|c|c|c|}
\hline \hline
 & Entity & Nature \\	
	\hline \hline
Level 7	& Internet &  Maximal Network \\ 
	\hline 
Level 6	& Network &Linked computers, printers, file servers  \\ 
	\hline 
Level 5	& Computer & Components  linked by bus, clock   \\ 
	\hline 
Level 4 & Components	& ALU, CPU, Memory, I/O devices
\\ 
	\hline 
Level 3	& Gates & Boolean logic: AND, OR, NOT   \\ 
	\hline 
Level 2	& Transistors & Binary ON/OFF function    \\ 
\hline 
Level 1	& Crystalline structure & Symmetry, Band Structure    \\ 
	\hline 
Level 0	& Ions, Carriers & Structure, Current Flow   \\ 
	\hline \hline
\end{tabular} 
\vspace{0.1in}
	
	\textbf{Table 2}:\label{Table2} \textit{Implementation hierarchy (schematic). This is the physical context within which upward emergence and downward causation takes place. Table 3 amplifies the lower levels.} 
	 
	 \vspace{0.1in}
\noindent The interface between the computer's software and hardware at level 4 is the \href{https://en.wikipedia.org/wiki/Instruction_set_architecture}{Instruction Set Architecture}  (ISA), which is the external (abstract) view of the computer chip (hardware).
 
 \subsection{Modularity} \label{sec:modularity}
  As discussed clearly by (\cite{Booch}:3-26), developing from  
  \cite{SimonHA}, in order to  obtain complex behaviour, each level $L_N$ should be made of  networks of modules at level $L_{N-1}$.  The key point is that we split a complex task up into simpler tasks, design units to handle the simpler tasks, and then knit them together in networks so that overall when taken together they handle the complex task. To do so, modules are characterised by (a)  Information Hiding, (b) Abstraction, (c) Interface Specification, and (d) being Modifiable. The situation is similar for physical modules and for logical modules, and the whole is done in a hierarchical way (sub-modules of modules do even simpler tasks). 
  
\begin{figure}[h]
	\centering
	\includegraphics[width=0.5\linewidth]{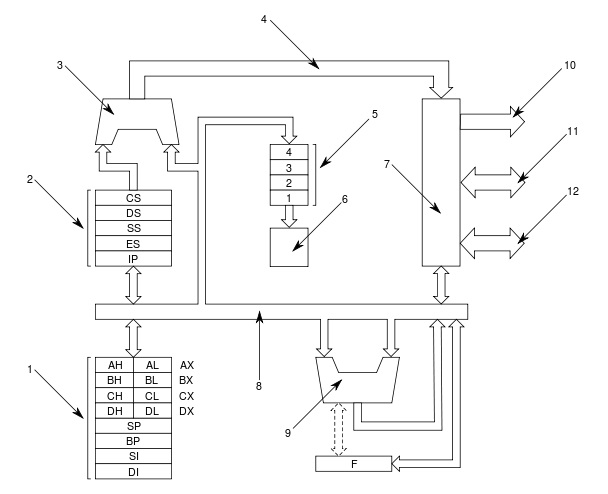}
	\caption{\textit{Simplified block diagram of Intel 8088.   1=main registers; 2=segment registers and IP; 3=address adder; 4=internal address bus; 5=instruction queue; 6=control unit; 7=bus interface; 8=internal databus; 9=ALU; 10/11/12=external address/data/control bus.} (Wikipedia, Intel 8086: open source)}
			\label{fig1}
		\end{figure}
  
  \paragraph{Interface Specification} The interface between each module and the rest of the system must be carefully specified, as in the case of the ISA: what control variables, parameters, and data will be sent to the module, and what will be sent out to other modules in the system from the module.
 
 \paragraph{Information Hiding} The  module's internal variables are hidden from outside view, all that is required is that the module does what it is required to do, as viewed from outside. The user does not mind what internal variables are chosen  and what logic is used.
 
 \paragraph{Abstraction} An abstraction is an external description of what the module does in a reliable way. This should include a name by which one can refer to the module, which indicates what it does (it should have a unique identifier). This can be seen in the case of the modules identified in Figure 1 (a representation of a specific case of Level 5 in Table 2).

 \paragraph{Modifiable}
 The previous points are directly related to multiple realisability (\S.\ref{sec:Multiple_Realis}) We can change the internal variables and even the internal logic of the module, as long as the externally viewed  functionality, as defined by the interface and abstraction, is maintained. The system as a whole only cares that the module does what it is designed to do, not how it does it (although some implementations are preferred over others if they are either faster, or use less resources).\\
  
 \noindent Figure 1 shows a typical set of physical modules that make up level 5 in the Implementation Hierarchy (Table 2).

\subsection{Multiple realisability} \label{sec:Multiple_Realis}

Multiple realisability is a key indicator that downward causation is taking place \cite{Auletta}, \cite{Ellis2016}, because it indicates that higher level needs are driving lower level structure and outcomes. It does not matter which lower level entities are chosen to carry out higher level needs as long as they do the job; the effective causal entity at the lower level is the \textit{equivalence class} at that level that does what is needed (\cite{Anthony_multiple}, \cite{Ellis2016},  \cite{Bickle_multiple}). This is a central feature of digital computers: both hardware and software can be realised in multiple ways. A modular structure plays a key role in facilitating this (you can alter the module logic and internal variables as long as the interface to the outside world is unchanged).
\begin{quote}
    \textbf{Module Significance:} \textit{Looked at this way, the central reason why modules are important is precisely that they allow multiple realisability to happen}.
\end{quote}

\paragraph{Hardware} There are many different microprocessors with different instruction sets that can be used in a digital computer while allowing running of the same high-level software \cite{Tannen}. A classic example is a Java Virtual Machine \cite{Java}, allowing standard Java to be run on any hardware. Wikipedia (\href{Java virtual machine}{``Java Virtual Machine''}) 
describes it thus: 
\begin{quote}``\textit{A Java virtual machine (JVM) is a virtual machine that enables a computer to run Java programs as well as programs written in other languages that are also compiled to Java bytecode. The JVM is detailed by a specification that formally describes what is required in a JVM implementation. Having a specification ensures interoperability of Java programs across different implementations so that program authors using the Java Development Kit (JDK) need not worry about idiosyncrasies of the underlying hardware platform}.'' \end{quote} 


\paragraph{Software} Any algorithm can be run by any properly defined computer language - of which there are a great many. A particular algorithm (e.g. Bubblesort, see \S\ref{sec:Bubblesort} below) can be loaded in any desired high level language (FORTRAN, BASIC, COBOL, LISP, PYTHON, C, etc.) and then via compilers and interpreters \cite{Aho} the same logic will  be realised in the Virtual Machine hierarchy (Table 1) in a different language at each level. A key point here is that algorithms go hand in hand with data structures \cite{Lafore}; both the software and the data have to be represented correctly at each level. 

\paragraph{The relation between hardware and software}
A further key point is that the boundary between what is done in hardware and what is done by software is fluid. Tanenbaum (\cite{Tannen}:8) states ``\textit{A central theme of this book is that hardware and software are logically equivalent''}. It is a matter of convenience - and performance speeds - how the split is made. Thus for example memory can be virtual (\cite{Tannen}:429-453) or physical (\cite{Tannen}:159-173).

\subsection{Black boxing not coarse graining}\label{sec:black_box}

As pointed out long ago by Ross Ashby \cite{Ashby}, when one is dealing with modules that perform logical operations, the appropriate method of going from a lower to a higher level is not coarse graining, it is \textit{black boxing}. One assembles lower level units  that perform specific logical operations $O_i$ in networks into an entity that  then comprise a next-higher level module performing operations $O_I$. Examples are how transistors can be connected to give modules that perform arithmetic and logic operations, including being decoders, multiplexers, flip-flops (the basis of memory), registers, and counters (\cite{Mellis}:44-58; \cite{Tannen}:136-164). A key feature here is multiple realizability, as discussed above: different circuits can give equivalent logical results, as e.g. in the case of the half-adder (\cite{Mellis}:45). Boolean algebra can be used to determine equivalent circuits (\cite{Tannen}:141-145). 

\section{Branching logic at all Levels}\label{sec:Branching_LOgica}

Causation is a contested word. However in the case of digital computers, it is quite clear it is taking place when the computer is used.
A program is loaded, and the operator runs the program. What happens at the physical level is determined by what program is loaded. Thus as mentioned above, we rely on a counterfactual view of causation \cite{Menzies}: if a different program is loaded, different flows of electrons will take place at the bottom level.\footnote{There are of course lower levels than Level 0 in Table 2, as described by the standard model of particle physics, which may in turn depend on even lower levels such as string theory/M theory. They are of no concern to us here.} This takes place by a flow of causation from the logical level to the physical level, whereby logical branching (\S\ref{sec:Branch_logic}), exemplified by Bubble Sort (\S\ref{sec:Bubblesort}), leads to physical branching (\S\ref{sec:Branch_physics}), with logical branching being in charge  (\S\ref{sec:Branch_relation}).

\subsection{Logical Branching}\label{sec:Branch_logic}

At user program level, one has logical branching of the form IF .. THEN .... ELSE, or   WHILE ... DO ...., or REPEAT UNTIL .....
This logic is chained down by compilers or interpreters to lower abstract machine levels, as discussed above (see Table 1). Given this logic, branching is governed by the relevant data.
At each level $M_I$ there is a language $L_I$ with a syntax that allows such branching; the details of the syntax varies at each level, and with the specific software implementation.

\subsection{Example: Bubblesort}\label{sec:Bubblesort}
To give a specific example of logic implemented in a computer, \textit{Bubblesort} is one of the simplest sorting procedures (\cite{Lafore}:79-89). In pseudocode, it can be written \href{https://en.wikipedia.org/wiki/Bubble_sort}{(Wikipedia)} as follows
\\

\vspace{5pt}
\hrule
\vspace{6pt}

\begin{quotation}
1	procedure bubbleSort ( A : list of sortable items )

2 \hspace{0.2in} n = length(A)

3 \hspace{0.2in}  repeat

4 \hspace{0.4in} newn = 0

5 \hspace{0.4in} for i = 1 to n-1 inclusive do

6 \hspace{0.6in} if A[i-1] $>$ A[i] then

7 \hspace{0.8in} swap(A[i-1], A[i])

8 \hspace{0.8in} newn = i

9 \hspace{0.6in} end if

10 \hspace{0.4in} end for

11 \hspace{0.4in} n = newn

12 \hspace{0.2in}  until n $\leq$ 1

13 end procedure

\end{quotation}

\vspace{5pt}
\hrule
\vspace{6pt}
Note the branching operations at lines 6 and 12 (there is always an implicit assumption ``IF  NOT, just carry on''). 
How this is implemented in $C^{++}$ is shown in \href{https://www.geeksforgeeks.org/cpp-program-for-bubble-sort/}{https://www.geeksforgeeks.org/cpp-program-for-bubble-sort/} which also gives links to implementation in Java and Python. For an assembly language version, see
\noindent \href{https://www.geeksforgeeks.org/8085-program-bubble-sort}{https://www.geeksforgeeks.org/8085-program-bubble-sort}. This variety of implementations of the same logic is of course a demonstration of multiple realisability.

Furthermore there are a variety of other sorting algorithms, e.g.  Mergesort, Quicksort, Heapsort, etc (\cite{Lafore}:89-108,279-294,315-364). They have been extensively compared for efficiency (\href{https://en.wikipedia.org/wiki/Sorting_algorithm}{\textit{Wikipedia: Sorting Algorithm}}). The point here is again that of multiple realisability and modularity: many programs need a sorting subroutine, so the logical need is simply an algorithm which will sort a list; the operational need is a module that will be efficient in terms of the type of list that will need to be sorted. It is that functional requirement that drives the designer's choice of a specific sorting algorithm in a particular program. Thus higher level need acts down to select lower level options from amongst a variety of choices (a classic example of adaptive selection).

\subsection{Physical Branching}\label{sec:Branch_physics}
Corresponding to the logical branching, physical branching takes place at each level in Table 2:\\

\textbf{At transistor level}, the branching is ON or OFF, depending on the gate voltage $V(t)$ (see (\ref{eq:threshold})).\\

\textbf{At gate level}, the branching is in terms of  truth values of basic logical operations: AND, NOT, NOR  which can then lead to branching in comparators, adders, decoders, etc. (\cite{Mellis}:37-58, \cite{Tannen}:135-164), with  outcomes depending on the data. For example, current flows take place in connections to the AND gate according to this logic, see the Wikipedia article on the \href{https://en.wikipedia.org/wiki/AND_gate}{AND Gate}. The physical structure has been shaped so as to implement this logic.\\

\textbf{At computer level}: branching occurs via CPU control of the basic instruction FETCH-EXECUTE cycle (\cite{Mellis}:75, \cite{Tannen}:173-202), with  activation of different instruction memory and data  memory locations and output modules via a bus (\cite{Tannen}:176-220) controlled by a clock. Branching of electron flows in the data bus  occurs according to the relevant data, as discussed in the Wikipedia article on the \href{https://en.wikipedia.org/wiki/Instruction_cycle}{Instruction Cycle}.\\

\textbf{At network level}: branching is for instance via requests represented by a Hypertext Transfer Protocol (\href{https://en.wikipedia.org/wiki/Hypertext_Transfer_Protocol}{HTTP}), sent between computers identified by  their Internet Protocol address (\href{https://en.wikipedia.org/wiki/IP_address}{IP address}), as for example sending a request to Google for information. The request is implemented by the transmission through the \href{https://en.wikipedia.org/wiki/Internet}{Internet} of electrons, photons, or electromagnetic waves, depending on the means of transmission,   according to the Internet Protocol Suite (TCP/IP).

\subsection{Relation}\label{sec:Branch_relation}
Logical branching governs physical branching, which is demonstrable by changing the program loaded. 
Outcomes alter, although the  physical structure involved (the computer hardware)  is identical.
This happens via compilers \cite{Aho} and interpreters  \cite{Tannen} that chain logic down to the machine code level. The logic of the algorithm \cite{Knuth}, for example Bubblesort (\S\ref{sec:Bubblesort}),  is preserved during the downward chaining  process, as it gets rewritten in different languages $L_I$ with different variables and syntax (Table 1).
The abstract becomes physical at the machine level, where digital logic is expressed in a timed sequence of electron flows that turn transistors ON or OFF (\cite{Tannen}:2-7, 231-326; \cite{Mellis}:75-80).

\section{Contextual branching at the transistor level}\label{sec:contextual_branch_Transistor}

Time dependent control signals at the machine level alter gate voltages $V(t)$ 
of integrated circuit transistors (see Figs. 2 and 3).
This changes a transistor from OFF to ON or vice versa, depending on a threshold level. In the  MOSFET case shown in Figs. 2 and 3, the logic is, \begin{equation}\label{eq:threshold}
\{IF\,\, V(t) >V_{threshold}\,\, THEN\,\, current\,\, flows\,\, ELSE\,\, not\}
\end{equation}
The question is how this branching logic relates to a unitary description at the microscopic level, 
which does not \textit{per se} allow branching (cf. \cite{Ellis_Koppel}). We discuss in turn, the interacting levels (\S\ref{sec:transistor_levels}); 
the disconnect between descriptions at different levels
(\S\ref{sec:disconnect}); 
and the downward links for branching causation (\S\ref{sec:down_links}). 

\subsection{The interacting levels}\label{sec:transistor_levels} To understand how the physical makeup of transistors can enable logical branching, one needs to 
look in more detail at the  physical hierarchical structuring at the transistor level (Table 3, in effect an expansion of the lower levels of Table 2).\\

\begin{tabular}{|c|c|c|c|}
	\hline 
	\hline 
		&  Level &Structure & Outcomes \\ 
	\hline 
	T5	&  gates & transistor combinations & Boolean Logic \\ 
	\hline 
	T4	& transistors & base, emitter, collector; carrier channels  & ON/OFF \\ 
	\hline 
	T3	& crystal structure & ions; symmetry, broken by impurities & phonons, band structure \\ 
	\hline  
	T2	& Electron population & densities, average flows, resistance &   electron diffusion, current \\
	\hline 
	T1	& Individual ions, electrons  & ion bonding; electron velocities, collisions & electron drift \\ 
	\hline 
	\hline 
\end{tabular} 

\vspace{0.1in}

\textbf{	Table 3:}\label{Table3} \textit{The physical (implementation hierarchy) at the transistor level.}

\paragraph{\textbf{Level T5: Gates}} Transistors linked via wires and resistors form the basic logical gates. Some of them can be made of a single integrated set of transistors.

\paragraph{\textbf{Level T4: Transistors}}Transistors are based in semiconductors such as silicon that have been doped with donor (type n) or acceptor (type p) impurities. Considering \textit{Field Effect Transistors} 
there is an (n,p,n) structure.
\begin{figure}[h]
	\centering
	\begin{minipage}[b]{0.4\textwidth}
		\includegraphics[width=\textwidth]{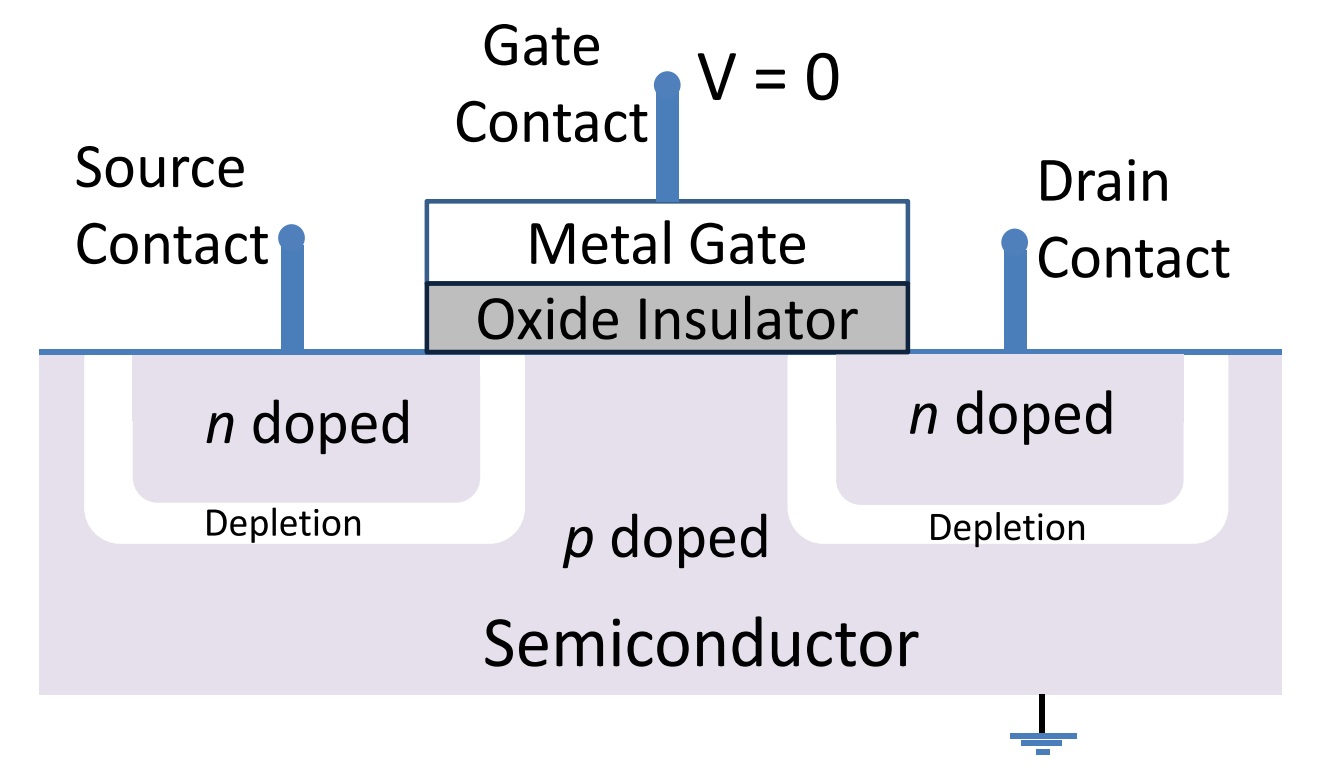}
		\caption{MOSFET with no bias applied to gate. The depletion layers separates the n-doped and p-doped regions, so no current flows. Source: \cite{Simon}:204. With Permission from OUP and the author}
		\label{fig2}
	\end{minipage}
		\hfill
	\begin{minipage}[b]{0.4\textwidth}
		\includegraphics[width=\textwidth]{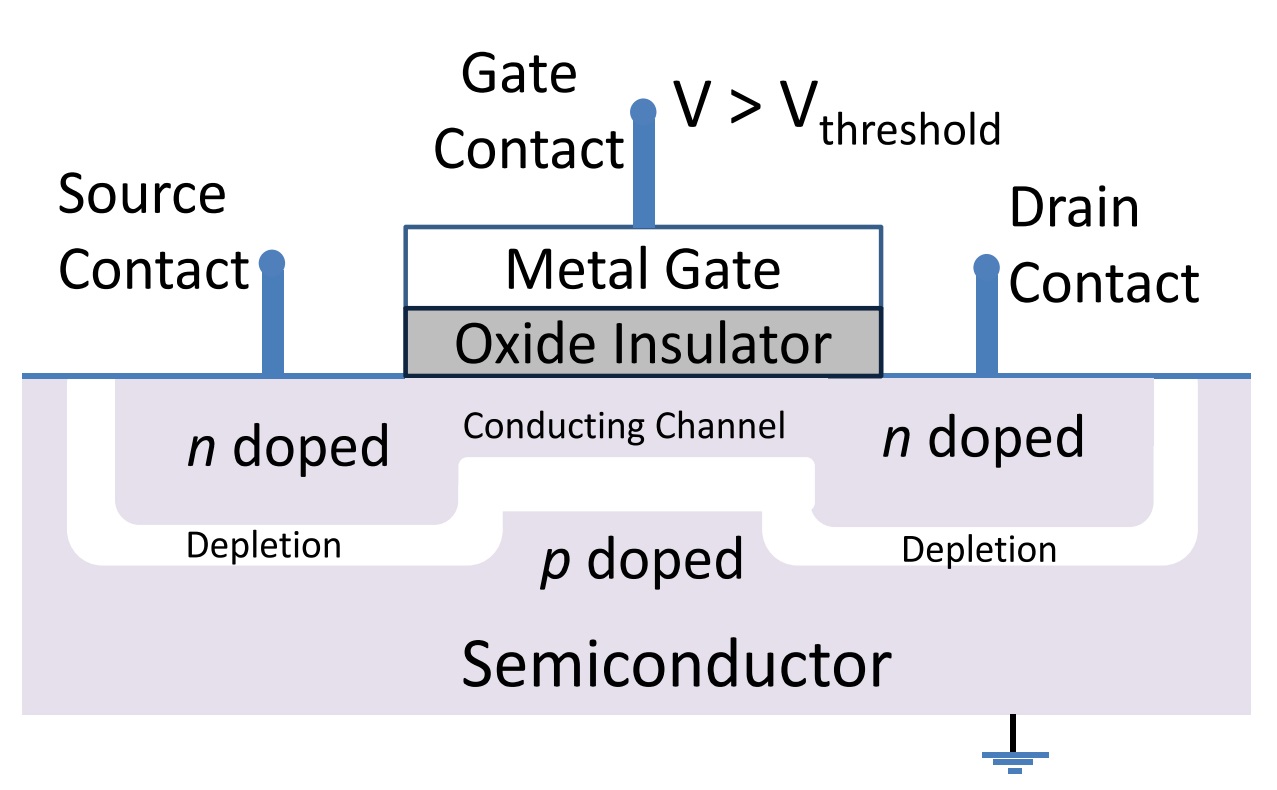}
		\caption{
			MOSFET with bias applied to gate. 
			The depletion layer shifts, so a continuous channel forms between the source and drain and current flows. Source: \cite{Simon}:205. With Permission from OUP and the author}
	\label{fig3}
	\end{minipage}
\end{figure}
 Fig.2 shows it when not conducting and Fig.3 when it is conducting. The gate is separated from the p-and n-doped semiconductor regions by an oxide insulator, making a capacitor. An applied voltage on the gate attracts electrons from the source and thereby opens a conducting channel between the source and drain, through alterations to the chemical potential and depletion region. 
 
\paragraph{\textbf{Level T3: Crystal structure, phonons, electronic bands}} The crystal structure with its particular symmetries and degrees of freedom gives rise to the various types of phonons and to the electronic band structure. Depending on the distance between the bands and their filling, one obtains the distinction between  conductors, insulators, and semi-conductors (\cite{Mellis}). 
In a semiconductor, conduction occurs only at nonzero temperature as electrons need to become excited from the valence band to the conduction band.

\paragraph{\textbf{Level T2: Electron population}}
Electron/carrier flow is due to (1) Diffusion (due to a density gradient), leading to depletion regions; (2) Drift due to an electric potential, leading to a current.  Resistance, characterised by a collision time, occurs due to collisions with impurities, phonons, and other electrons.
Electron conduction is time asymmetric because of interaction with a heat bath. Due to its dissipative nature, electron conduction is not a Hamiltonian process. 

\paragraph{\textbf{Level T1: Ions and electrons}}
The description on the most microscopic level is based on a Hamiltonian for the ions and electrons  (\cite{Philips}:16):
\begin{eqnarray}
H &=& 
-\sum _i\frac  {\hbar ^2}{2M_i}\nabla _{{\mathbf  {R}}_i}^2
-\sum _{i}\frac {\hbar ^{2}}{2m_e}\nabla _{\mathbf {r} _{i}}^{2}
+ \sum _i\sum _{j>i}\frac  {Z_{i}Z_{j}e^{2}}{4\pi \epsilon _{0}\left|{\mathbf  {R}}_{i}-{\mathbf  {R}}_{j}\right|} \nonumber\\
&& -\sum _{i}\sum _{j}{\frac  {Z_{i}e^{2}}{4\pi \epsilon _{0}\left|{\mathbf  {R}}_{i}-{\mathbf  {r}}_{j}\right|}}
+ \sum _{i}\sum _{{j>i}}{\frac  {e^{2}}{4\pi \epsilon _{0}\left|{\mathbf  {r}}_{i}-{\mathbf  {r}}_{j}\right|}}\label{eq:Hamiltonian}
\end{eqnarray}
 
 However, in order to derive from this Hamiltonian the actual lattice structure and electron charge distribution in the transistor, a series of approximations is required: 

A:  First, the electrons are separated into conduction band electrons (essentially unbound and so free to move) and valence band electrons (closely bound to ions and so localised). This transforms the Hamiltonian (\ref{eq:Hamiltonian}) to the form, 
\begin{equation}
H = T_i + T_e + V_{ii} + V_{ee} + V_{ei} + E_{core}
\end{equation}

B: Next, the Born-Oppenheimer (adiabatic) approximation is used, which assumes that the electrons are at all times in equilibrium with the positions of the ions, see for instance \cite{schwablqm}, Ch.~15.
To this purpose, the wave function is factorized into an electron part $\Psi_{e}(\textbf{r},\textbf{R})$ for given positions of the ions, and an ion part $\Phi(\textbf{R})$,
\begin{eqnarray}
\Psi(r,\textbf{R}) =  \Phi(\textbf{R}) \Psi_{e}(\textbf{r},\textbf{R}) \, ,
\end{eqnarray}
leading to the electron equation 
\begin{eqnarray} \label{electronequation}
(T_e+V_{ee}+V_{ei}) \Psi_{e}(\textbf{r},\textbf{R}) = E_{e}(\textbf{R})\Psi_{e}(\textbf{r,R})
\end{eqnarray}
 and the ion equation  
\begin{eqnarray}\label{ionequation}
(T_i+V_{ii}+E_{core}+E_{e}(\textbf{R})) \Phi(\textbf{R}) =  E\Phi(\textbf{R})  \, .
\end{eqnarray}

C: The electron equation (\ref{electronequation}) is used to obtain the electronic band structure. Since the band structure is based on a picture of non-interacting electrons, the interaction term $V_{ee}$ must be dropped so that the electron Hamiltonian becomes a sum of one-particle Hamiltonians. 
Given a periodic lattice, the solutions are Bloch waves, giving the energy bands $E_n(\textbf{k})$, which are the lattice analog of free particle motion (\cite{Philips}:22).

D: Electron-lattice interactions occur via phonons (\cite{Simon}:82-84, 90-95). The ion equation (\ref{ionequation}) can be used to derive the phonon modes of the ions by making several further approximations (localized ions, harmonic expansion of the energy around the equilibrium ion positions).  The normal modes of the resulting harmonic model determine the dispersion relations of phonons.

E: Finally, in order to model electron-phonon interactions explicitly, a quantum field theoretical formalism is required that is based on  creation and annihilation operators for electrons and phonons, see for instance \cite{solyom2}, Ch. 23. The dispersion relations of electrons and phonons obtained by the above-mentioned methods determine the electron and phonon propagators, but the interaction term requires a separate evaluation of the cross section for the scattering of electrons via the emission or absorption of a phonon. 

\paragraph{Doping Effects}
Doping with impurities adds electrons to the conduction band (donor, or n-doping) or holes to the valence band (receptor, or p-doping) (\cite{Simon}:187-194).

\paragraph{p-n and n-p Junction effects} The junctions between n-doped and p-doped regions are the key feature of transistors (\cite{Mellis}:14-20; \cite{Simon}:199-203). Electron diffusion at such junctions leads to depletion regions, stabilised by induced electric fields (\cite{Mellis}: Fig.1.9). 
Transistors are created by suitably shaped such junctions (\cite{Simon}:203-205).  

\paragraph{Electric field effects} We do not want to model the dynamics of a transition between ON/OFF states in the transistor, but rather the steady state situation when it is in  one or other of these states (conducting or not) according to whether the applied voltage is above a threshold or not. This means that electric field effects must be modelled.

To do this, the gate  voltage  $V(t)$ 
 must be added to the model. This leads to a potential energy term  in the Hamiltonian of the electrons:
 \begin{equation}\label{HV} H_V(t)=\sum_i e V(\textbf{r}_i,t)\end{equation}
 where the Level $T_4$ variable $V(t)$ determines the  Level $T_1$ variables $V(\textbf{r}_i,t)$ in a downward way. It leads  to a displacement  
 of the electrons until a new equilibrium is reached where the electrical field created by the modified charge distribution cancels the electrical field due to the gate voltage. In order to calculate this new equilibrium, a self-consistent calculation based on the charge density due to doping, gate potential, and thermal excitation must be performed. 

This alters the band structure (\cite{Mellis}:19,  \cite{Simon}: Fig.18.6) and thereby either creates a conduction channel by changing the depletion zone, or not,  according to the bias  voltage applied (compare Figs. 2 and 3 above). 
 In the case of the ON state, an additional electric field between the two contacts leads to electron conduction, which will be discussed in  more detail further below.  
 
\subsection{The disconnect}\label{sec:disconnect}
The previous description has shown that even though one starts with the electron-ion Hamiltonian, one makes subsequently a series of approximations that effectively replace the initial model by other models. These other models involve classical elements and elements from statistical physics, in addition to Hamiltonians. It is this combination of elements that enables the switching between the ON and OFF state of the transistor.\\

(1) \textbf{Level $T_1$}: 
The Born-Oppenheimer approximation is based on the idea that ions are localized objects, neglecting entanglement between ions and electrons. This is an element from classical physics, and it is essential for obtaining the lattice structure, which breaks the symmetry of the underlying Hamiltonian \cite{primas,chibaro}. A purely quantum-mechanical calculation cannot give such a symmetry-broken ground state. However, this symmetry-broken state is the basis for deriving electronic band structure and phonons (cf \cite{Anderson}, \cite{Philips}:1-2). This means that the new Hamiltonians that describe electrons in bands and quasiparticles called phonons are based on a new,  phenomenological theory that includes essential classical ingredients.\\

(2) \textbf{Level $T_2$}: Electron conduction on the most microscopic level is typically described using the Boltzmann equation (\cite{Philips}:179-187), which includes collision rates as basic ingredient. This means that a model with localized electrons and stochastic transitions is used, neither of which feature in the initial Hamiltonian. This description is therefore not derived from a microscopic Hamiltonian, but it is the appropriate form of physical dynamics that applies at that level: it has its own logic that is not fully determined by the underlying quantum physics. \\

(3)  \textbf{Level $T_3$}: The derivation of the phonon dispersion relations involves the quantization of classical harmonic oscillations, which in turn are obtained by approximating the ion equation from the Born-Oppenheimer approximation by an harmonic expansion around the ion equilibrium positions. This means that one is using a classical intermediate model,  the low-lying excitations of which are quantized by hand. This leads to a new Hamiltonian, which again is not really derived from the supposedly fundamental Hamiltonian for the electrons and ions. \\

(4) \textbf{Level $T_4$}: In order to calculate the charge in the depletion region, one performs a classical electrostatics calculation based on the electron density in the conduction band obtained from statistical physics calculations of the thermally excited electrons. Again, this is not derived directly from the base Hamiltonian, because contextual information about the structure of the transistor, the doping, the gate voltage, and the temperature need to be taken into account. \\

Overall there is no continuous derivation of the relevant properties from the Hamiltonian \ref{eq:Hamiltonian}. There are rather a series of empirically adequate phenomenological models building on each other but not derivable from them. As stated by Leggett
\cite{Leggett1992},
\begin{quote}
\textit{``No significant advance in the theory of matter in bulk has ever come about
through derivation from microscopic principles. (...) I would confidently argue
further that it is in principle and forever impossible to carry out such a derivation. (...) The so-called derivations of the results of solid state physics from
microscopic principles alone are almost all bogus, if `derivation' is meant to have anything like its usual sense. Consider as elementary a principle as Ohm’s
law. As far as I know, no-one has ever come even remotely within reach of deriving Ohm's law from microscopic principles without a whole host of auxiliary
assumptions (`physical approximations'), which one almost certainly would not
have thought of making unless one knew in advance the result one wanted to
get, (and some of which may be regarded as essentially begging the question).''} \\

\textit{``This situation is fairly typical: once you have reason to believe that a certain
kind of model or theory will actually work at the macroscopic or intermediate
level, then it is sometimes possible to show that you can `derive' it from microscopic theory, in the sense that you may be able to find the auxiliary assumptions
or approximations you have to make to lead to the result you want. But you
can practically never justify these auxiliary assumptions, and the whole process is highly dangerous anyway: very often you find that what you thought you
had `proved' comes unstuck experimentally (for instance, you `prove' Ohm's law
quite generally only to discover that superconductors don't obey it) and when
you go back to your proof you discover as often as not that you had implicitly
slipped in an assumption that begs the whole question. ''}
\end{quote}
\begin{quote}
\textit{``Incidentally, as psychological fact, it does occasionally happen that one is led to a new model
by a microscopic calculation. But in that case one certainly doesn't believe the
model because of the calculation: on the contrary, in my experience at least one
disbelieves or distrusts the calculation unless and until one has a flash of insight
and sees the result in terms of a model. I claim then that the important advances in macroscopic physics come essentially in the construction of models at an
intermediate or macroscopic level, and that these are logically (and psychologically) independent of microscopic physics.''}
\end{quote}
Such approximations and models are introduced at each step up in the representation of the implementation hierarchy. At each stage, symmetry breaking takes place in accord with Anderson's discussion \cite{Anderson}, as emphasized particularly by Phillips \cite{Philips}.

\subsection{Branching causation: The downwards links}\label{sec:down_links}
The depletion region grows or shrinks depending on the applied voltage. This cannot be described in a Hamiltonian way, because it is based in diffusion processes. Furthermore the current flow is determined by collisions with impurities and phonons, as characterised by a scattering time $\tau$ (a micro variable) associated with the resistance of the material (a macro variable) (\cite{Simon}:19). In both cases one has irreversible dynamics arising from the underlying unitary dynamics in the context of the transistor structure, and the issue of the arrow of time occurs. Ultimately this is resolved by Contextual Wavefunction Collapse, where heat baths in contact with the external environment determine the local arrow of time from the cosmological direction of time \cite{Drossel}. \\

This opening or closing of the conduction channel happens in accord with the  instructions at the binary logic level, via the basic computer functioning cycle (fetch half cycle, execute half cycle) (\cite{Mellis}:75-76; \cite{Tannen}:54-58) executed via  modules such as depicted in Figure 1. This happens according to the logic of the algorithm being implemented (see Table 1).\\

\textbf{The bottom line} The entire design of the transistor is chosen such that it can act as an ON/OFF switch triggered by the gate voltage: 
\begin{quote}
    \textbf{The crucial dynamic}  \textit{When a time dependent bias voltage $V(t)$ is applied to a transistor gate, it alters the underlying Hamiltonian through the time-dependent potential term $H_V(t)$, see (\ref{HV}), so usual existence and uniqueness theorems do not apply: outcomes are not determined by initial data.\footnote{See the Appendix of \cite{Ellis_Koppel}.}
    Furthermore the dynamic at work is anyway not just about changing a Hamiltonian, but also about the macroscopic world, which has classical features that arise from ongoing localization and symmetry breaking due to interaction with heat baths, which affect what happens at the transistor level. This changes outcomes in a way determined by the  potential $V(t)$, as expressed in  (\ref{eq:threshold}). 
This is how the logic represented in the abstract structure of the computer program, expressed in machine code,  controls the underlying physics at the electron level; the electron dynamics is no longer unitary}. 
\end{quote}
Outcomes are determined by the time dependent function $V(t)$ determined by the machine code (Level $M_1$ in Table 1), and applied in the context of the specific detailed structure of the transistor concerned (Wikipedia, \href{https://en.wikipedia.org/wiki/Transistor}{Transistor}; cf. Figs. 2,3). We represent that structure in a phenomenological way rather than by detailed constraint equations in the Hamiltonian formulation. 

\section{The nature of variables}\label{sec:variables_nature}
Downward causation refers to causal influence of higher level variables over lower level variables. This requires that the relevant higher level variables either have 
emerged strongly
 (\S\ref{sec:variables_High}),  or are of a character that does not emerge from the lower levels because they are of a quite different kind: they are abstract rather than physical variables (\S\ref{sec:variables_High_logic}). In the latter case, the logical-physical interface is crucial (\S\ref{sec:Logical_Physical}). 
 It requires then that 
 in each case the higher level 
 variables 
 reach down to constrain 
 lower level variables, as discussed in the previous sections.

\subsection{Irreducible Higher Level Variables: Physical}\label{sec:variables_High} 

Intrinsically higher level variables
cannot be obtained by coarse graining lower level variables: they are not `nothing but' a summation of lower level effects \cite{Anderson}. A key distinction is between global versus local variables: the former are characterised by higher level structure, which needs to be described at its own level; its essential causal features do not exist at any lower level.\\


In the case of the computer, the transistor structure as a whole (cf Figures 2 and 3) is a Level $T_4$ feature, determining how lower level entities interact with each other. Lattice vibrations (\cite{Philips}:171-172)
are a property of the crystal as a whole, which cannot be described or understood at any lower level. Thus \href{https://en.wikipedia.org/wiki/Phonon}{phonons} are a collective property arising from  the macroscopic crystal structure 
(\cite{Philips}:169-172), \cite{Simon}. The \href{https://en.wikipedia.org/wiki/Depletion_region}{depletion zone} is a property of the specific transistor construction \cite{Mellis}, so is again a higher level variable. In each case these higher level features are as a matter of fact structural features that only occur at that level. Furthermore, this is not just an issue of description: the materials have to be assembled in a specific way to make them come into existence. Doing so is a costly enterprise in terms of both design and manufacture \cite{Mellis}. They exist ontologically whether we know about them or not.

\subsection{Irreducible Higher Level Variables: Logical}\label{sec:variables_High_logic}
A computer program as a whole \cite{Abelson} is an irreducible higher level entity.  The \href{https://en.wikipedia.org/wiki/Scope_(computer_science)}{scope} of a variable is a key feature of a program; global variables are higher level quantities. Passing global parameters or variables to subroutines in a computer program changes logical constraints at the lower level.\\

All this is a logical structure, described in the specification manual for the language $L_i$ at some level $i$ in table 1. 
Because it is an abstract entity, it is simply of a quite different nature than the physical structures on which it supervenes. It cannot be derived by coarse graining the underlying physical structure: to attempt to do so would be a category mistake, for example you cannot determine the Bubble Sort algorithm (\S\ref{sec:Bubblesort}) by coarse-graining the physical structure and electron flows at any level in the implementation hierarchy (you might possibly be able to logically deduce it from such information; that is a completely different process, precisely because it is logical in nature). Nevertheless that abstract relation is what determines the physical outcomes of a sorted list on a computer screen or a physical printout of a list. This is discussed further in Section \ref{sec:causal_power} below. 

\subsection{The logical-physical interface}\label{sec:Logical_Physical}

\href{https://en.wikipedia.org/wiki/Variable_(computer_science)}{Logical variables} - which are defined in digital computers at each level of the virtual machine hierarchy (Table 1) - cannot be derived from physical variables because they are simply of a completely different nature. Rather they can be \textit{represented} by physical variables - which cannot possibly be a bottom-up process, because the very concept of logical variables is not a physics concept. \\

They form a \href{https://en.wikipedia.org/wiki/Formal_language}{formal language}: that is,  \textit{``they consist of words whose letters are taken from an alphabet and are well-formed according to a specific set of rules''} (Wikipedia). 
 Those rules are an abstract structure \cite{Abelson}, formalised in the language specification, see for example \cite{Java} for the Java language specification. The language is represented physically in the computer by a map to physical variables. Thus we have a map  ${\cal S}: V_A \rightarrow S_i$  between words  $V_A$  and a set of physical 
 entities $S_i$.  To be useful, the map must preserve the rules of the language; that is what is achieved by the way computers are structured \cite{Tannen}.   Then the physical operation of the system, via flows of electrons in transistors, implements that formal set of rules (for example, Bubblesort as in  \S\ref{sec:Bubblesort}, written in terms of a specific computer language such as Java \cite{Java}). \\
 
 A key point is the following: computing famously costs energy according to \href{https://en.wikipedia.org/wiki/Landauer27s_principle}{Landauer's principle}. However it should be clear from the above that one cannot determine the outcome of a computation by application of any kind of physical energy minimisation or entropy maximisation principle. The computations proceed on the basis of the logic of the algorithm as realised in the chosen language, which has no direct link to energy utilisation.  Energy limitations certainly place limits on what kind of computation can be done in practice by limiting the number of operations that can be performed per second, as for example in \href{https://en.bitcoin.it/wiki/Mining}{BitCoin mining}, but no principle based in physical energy usage can determine logical outcomes due to implementation of algorithms, as discussed above. There are two specific  points to  make here. Firstly, the algorithm may involve calculation that is logically equivalent to an energy minimisation process, for example optimisation of a function, or Bayesian probability estimation (thus energy minimization in an information theoretic sense \cite{Tutorial}). That is not the same as saying that the physical implementation of the algorithm minimizes energy. That depends on the number of computational steps involved in the algorithm, so physical energy use is minimized by algorithm and program optimization: that is, finding another algorithm and program which will do the same logical job using fewer computational steps. Studying how to do this is a key aspect of algorithm and program design \cite{Knuth}.

\section{Synchronic and Diachronic Supervenience}\label{sec:supervene}
It is claimed by some that downward causation is not possible because of supervenience,  see \cite{GibbHendryLancaster} for an extensive discussion. 
\textit{Synchronic supervenience} is the idea that any specific lower level state $L_1(t_0)$ at time $t_0$ leads to a unique higher level state $H_1(t_0)$ at that time. 
\textit{Diachronic supervenience} is the idea that any specific lower level state $L_1(t_0)$ at time $t_0$ leads to a unique higher level state $H_1(t_1)$ at a later time $t_1$. Its justification is the combination of synchronic supervenience at both times $t_0$ and time $t_1$ with the alleged unitary dynamics at the lower level: i.e the assumption that the initial lower level state $L_1(t_0)$ uniquely determines the later lower level state $L_1(t_1)$.
However we have seen above that the latter assumption is not generically true: lower level branching dynamics can occur due to higher level variables changing constraints at the lower levels (a simple example is a pendulum of varying length, see the Appendix of  \cite{Ellis_Koppel} for details). Here are some examples of the possibilities (\S\ref{sec:supervene_programs}-\S\ref{sec:supervene_unpredictable}).  

\subsection{Fixed Programs and data}\label{sec:supervene_programs}
If the program ${\cal P}$ has fixed initial data ${\cal D}$ that is loaded with the program, so that at time $t_1$ both ${\cal P}$ and ${\cal D}$ are in memory, then diachronic supervenience will hold from that time on: the computer will  run according to that program  and data, and the outcome is uniquely determined at each time $t > t_1$ (provided the program halts: we will not deal with that issue here).  
But how did the program and data get there? They were not there at an earlier time $t_0$. The initial state $S(t_0)$ when the program and data have not been loaded cannot predict the later state $S(t_1)$ 
when
they are in memory.
  Thus  diachronic supervenience  will not be true from time $t_0$ to time $t_1$.

\subsection{Interactive programs}\label{sec:supervene_interact}
Suppose data is input on an ongoing basis by  sensors, for example in the case of aircraft automatic landing system. The data ${\cal D}(t)$ used such as height, wind speed and direction, direction to the airport, position of other aircraft, and so on  is updated  on the basis of radar,  GPS, and other readings. These  high level variables are not predictable from the microphysics \textit{inter alia} because of atmospheric turbulence (which can exhibit chaotic dynamics). Thus they are not uniquely predictable from the earlier data. Thus diachronic emergence does not hold.
The training of artificial neural nets (ANNs) is another example. Weights of links between nodes are determined by backpropagation algorithms and  supervised  learning methods, which depend on the order in which training examples are used and the specific data set chosen. These are not predictable on the basis of the data initially available to the computer.

\subsection{Adaptive programs}\label{sec:supervene_adaptive}
Programs that learn, for instance evolutionary computation based in genetic algorithms \cite{genetic}  go a step further: the program ${\cal P}(t)$ is a function of time, and is optimised as execution proceeds. In such programs, a random number is chosen to seed the selection process, 
which can be based for example on the precise time when program execution started, and hence is unpredictable from both a microphysics and macroscopic viewpoint. 

\subsection{Unpredictable input}\label{sec:supervene_unpredictable}
One can choose to run programs where program branch points are determined by quantum uncertainty. Thus one can have data specified by a sensor detecting particles emitted by decay of a radioactive atom, in which case the detection times and consequent program branchings are unpredictable even in principle because of the foundational nature of quantum uncertainty \cite{Ghirardi}. Diachronic supervenience is impossible in this case because the data ${\cal D}(t)$ cannot be known in advance of the quantum detection event.

\subsection{Evolutionary development}\label{sec:evolution}
Finally how did digital computers on the one hand, and transistors/integrated circuits on the other, come into existence? They did not exist 100 years ago, now they do. 
They came into being via a evolutionary process of creative  vision of possibilities leading to adaptive selection of technology and associated concepts, 
for example the idea of the transistor, its detailed development into a variety of implementations, and the  development of the complex manufacturing methods for VLSI chips (\cite{Mellis}:36-37).
\subsection{The larger context} A computer is not a closed system. It is an open system and therefore a Hamiltonian description fails anyway. It is open in many respects: new bits from the computer program, it needs electrical power (which can fail), is in contact with heat baths, has finite temperature and emits therefore photons, and so on. Furthermore cosmic rays cause a non-zero error rate in computers \cite{Cosmic_rays}, and they are in principle unpredictable because their emission is a quantum event. The basis for diachronic emergence is not there.

\subsection{Causal Exclusion} Finally, what about the problem of causal exclusion?. As stated in Section 6.2 of \textit{The Stanford Encyclopedia of Philosophy} \cite{Stanford},
\begin{quote}
    ``\textit{Whenever any mental (functional) property M is instantiated, it will be realized by some particular physical property P. This physical property is unproblematically relevant to producing various behavioral effects. But then what causal work is left for M to do? It seems to be causally idle, “excluded” by the work of P}.
\end{quote}
What this omits of course is how has it occurred that the property M has come to be instantiated by the physical property P. The answer, as shown above, is by downward causation. This is what occurs in the case of algorithms implemented via computer programs, where the algorithm is the functional property M, and the physical property could be either at level T1 or T2 or T4 or T5 in Table 3. There is no other way that the Bubblesort algorithm (Section \ref{sec:Bubblesort}) could have been realised at any of these levels. This logical structure could not just have appeared by some random process at these levels (the probabilities of the precise ordered set of symbols needed at each level appearing  randomly is far too low) - or by the deterministic operation of the laws of physics alone at any of these levels, for there is no law of physics that necessarily produces the logic of Bubblesort (cf. \cite{Ellis2005}).

\section{Conclusion}\label{sec:conclude}

 This paper clarifies how the logical structure of computer programs, implementing suitable algorithms,  controls the flow of electrons at the transistor/gate level in digital computers. To conclude, we discuss in turn in this section, how abstract entities can have causal powers (\S\ref{sec:causal_power}), and hence physics \textit{per se} is not causally complete  (\S\ref{sec;Causal_completeness}); the difference between the argument presented here and the case of biology (\S\ref{sec:difference_biology}); the outcome that all levels are equally real (\S\ref{sec:levels_all_real}); and the social connection (\S\ref{sec:social_connection}), 
 
 \subsection{Abstract entities have causal powers}\label{sec:causal_power}
 Digital computers provide a clear example of \textit{downward causation} from  algorithms and data to the relevant underlying physical level (electron flows in semiconductor materials); this is what enables the dual pair of computer programs and their associated data to have genuine causal power - as they undoubtedly do \cite{McCormack_algorithms} - through a computational process. The nature of such a process is described by Abelson and Sussman as follows (\cite{Abelson}:p.1):
 \begin{quotation}
 ``\textit{Computational processes are abstract beings that inhabit computers. As they evolve, processes manipulate other abstract things called data. The evolution of a process is directed by a pattern of rules called a program. In effect, we conjure spirits of the computer with our spells. A computational process is indeed much like a sorcerer's idea of a spirit. It cannot be seen or touched. It is not composed of matter at all. However it is very real. It can perform intellectual work. It can answer questions. It can affect the world by disbursing money at a bank or by controlling a robot arm in a factory}''. 
 \end{quotation}
 Thus they are examples showing that a simple materialist position - the idea that only physical entities can have causal powers - cannot be correct:
\begin{quote}
 \textbf{Abstract entities have causal powers} \textit{ It is clear from this discussion that (1) algorithms, (2) computer programs, and (3) data - all abstract entities - have causal powers because they alter physical outcomes in a real world social context}.
   \end{quote} 
   An objection that can be raised to this  discussion is the following:\footnote{We thank an anonymous referee for this comment} we claim that diachronic supervenience fails because the alleged unitary dynamics at the lower level does  not hold, and lower branching dynamics can occur  due to higher level variables changing constraints at the lower levels. But there is no reason to suppose that the responsible higher level variables can only be the ``abstract'' entities from the logical hierarchy (which is the necessary intermediate assumption, in order to establish causal powers to abstract entities). The provided examples show clearly that computers are open systems, receiving extra inputs along the way that disrupt diachronic supervenience. But we can coherently (and plausibly) assume that every such input data is at the same time accompanied by analogous physical entities and processes that occur at every level of the physical hierarchy. \\

The response is that this is indeed what happens. This is nothing but a restatement of Noble's ``Principle of Biological Relativity'' \cite{Noble_2012} applied to the case of digital computers.
The key point is that the physical hierarchy is driven by the logical hierarchy, which determines what sequence of physical processes will in fact take place. This can be demonstrated by altering the algorithm, entering it as an altered code in terms of some high level languages (which is where the logic gets represented in physical terms), and then checking that different flows of electrons take place at the transistor level as a consequence. This is why the levels in the logical hierarchy (Table 2) are characterised as ``virtual machines'', as explained in depth by Tanenbaum \cite{Tannen}: they can be treated as if they are physical entities, even though they are abstract. This is not just a question of epistemology: each level is characterised by a set of symbols and associated syntax that as a matter of fact determine causation at that level (try putting a bracket in the wrong place in Latex, and see what happens!) The underlying electron flows are instantiations of that syntax.\\

\textbf{The bottom line} If  you use a sorting algorithm on a list of entities, the algorithm is abstract as is the list, but both are represented in ASCII code at a high level and binary digits at a low level. At an abstract level, the algorithm will result in an unsorted list of entities  $\{a_I\}$ becoming a sorted list; at the physical level, this will result in a representation of this list on a screen or on a printed output also being in sorted order. The algorithm is the causal agent; changing a logical switch will result in both the logical output and the physical output being in inverse order. This takes place within the logical constraints and physical constraints on the system,  but is an active combination of logical and physical processes for which the word `causation' is appropriate; see Section 6 of \cite{Ellis_Koppel} where deductive causation is discussed.\\

What of the suggestion that just remaining in the physical realm seems \textit{ontologically} adequate to account for the causal networks of the relevant parts of the computing world? The different claim that the series of higher level phenomenological models cannot be reduced to (or obtained by) the microphysics itself becomes then an \textit{epistemological} issue, having to do with our limitations as beings in the world trying to do science?\\

The response is that one is dealing here with the crucial issue of symbolism and the causal power of symbols (\S \ref{sec:Logical_Physical}) in the context of an automated computational system. Once a symbolic system is set up and the logical laws of interaction between symbols have been specified (in our case, the semantics of the programming language at any level in Table 1), and that semantic system has been realised via the compilers and interpreters loaded in the system, the computer proceeds to carry out mechanical computations based  on the logic of the algorithms incorporated in the program. No human consciousness or interpretation of the symbols intervenes in the course of that computation: this is the essence of AI, it is a symbolically driven machine process. Epistemology comes into play in interpreting the output (what do the symbols represent?) but not in the mechanics of the machine process - which is controlled by abstract logic. The implementation hierarchy is controlled by the logical hierarchy, and that is a fact about how digital computers work, irrespective of whether we know how this takes place or not (after tall, most computer users have no idea what is going on under the hood: the system just does what they tell it to do).  
   
   \subsection{Causal Completeness}\label{sec;Causal_completeness}
 This furthermore shows that consideration of digital computers gives a concrete example of how the underlying  physics 
 \textit{per se} is not causally complete, as is often alleged. Rather,   
 \begin{quote}
     \textbf{Causal completeness:} \textit{In the real world, it is only the combination of physics with its logical, social, psychological, and engineering contexts (which includes the values guiding policy) that can be causally complete, because it is this whole that determines what computer programs will be written and what data utilised, hence what electron flows will take place in integrated circuits, as per the discussion in this paper.}
  \end{quote}
As a specific example: the amount of money that can be dispersed to you from an ATM will be limited by an agreement you have reached with your bank. The program used to control the ATM will take into account the existence of such limits, and the specific amount you are able to take out in a given time period will be limited by a logical AND operation linking this agreed amount to the amount of money in your account. Thus these abstract variables will control electron flows in both the bank computers and the ATM dispenser mechanism.
Every relevant abstract variable has physical counterparts; in other words, it’s realized by some physical properties on some relevant physical substrate. For example, the stored abstract variable encoding the information about the available amount of money in my account possibly corresponds to, say, a specific polarization state of a ferroelectric material inside some distant memory device.

This poses problems for diachronic supervenience, but does it indicate  that strictly speaking it is not the abstract variable itself (here: the amount of money in my account) that controls the possible lower level branching dynamics; rather it is the physical properties of the memory unit causally responsible for the outcome dynamics? Why not just restrict ourselves to the claim that algorithms, computer programs, and data just express/capture/(or what have you) constrains possible physical outcomes, instead of going for the stronger claim that they “have causal powers because they alter physical outcomes”?\footnote{We thank an anonymous referee for this comment.}  Consider a well-known example (originally due to M. Lange) that a granny cannot evenly allocate to her 3 grandchildren all the 20 candies in her pocket; the reason being that 20 cannot be exactly divided by 3.\\

Now such mathematical facts constrain the relevant physical situations we may consider, as do physical constraints such as energy conservation. But it is a fact that if the limit $L$ to the amount of money I can draw is altered, the flows of electrons will be changed by that alteration to my  agreement with the bank (an abstract relation expressed in logical terms). It is not a result of a logical limitation (it can involve an agreed overdraft so negative quantities are allowed!). The causal feature is an assessment by the bank of the likelihood of my paying any overdraft, and that economic process has physical consequences. The word `cause' is indeed appropriate because of counterfactual reasoning.\\

As regards the granny mentioned above, while logical issues constrain her choice (20 is not divisible by 3), what occurs is not not merely due to a logical constraint. The same kind of issue may hold
for her as in the case of the bank: maybe she likes one grandchild less than the others; or
thinks that the small one can eat less candy; or one has sensitive teeth and should not have much sugar; or she wants to reward them according how well they have done at school, or has some other reason in
terms of values or preferences to choose an uneven distribution.
It can be other considerations than logical, and even values,  that determine what happens.
 
\subsection{The difference from biology}\label{sec:difference_biology} 
In the case of biology, much of the arguments about the effectiveness of top down causation, able to guide dynamic outcomes at the underlying electronic level, is the same as presented here \cite{Ellis_Koppel}. However there is a key difference.\\

Molecular biology is based in the 3-dimensional conformational  structure of biomolecules such as nucleic acids and proteins \cite{Watson}. 
Signalling molecules \cite{Berridge} convey messages about biological activity needed to meet higher level goals \cite{Noble_2012}, and cause a change in the conformation of other biomolecules, i.e., in distances between nuclei in  proteins or DNA \cite{Karplus}, thereby altering  the Hamiltonian governing the physics in a time dependent way. 
This is the mechanism whereby higher level biological needs can alter the outcomes at the ion/electron level (for full discussion, see  \cite{Ellis_Koppel}). This contrasts with the case considered 
here, where the distances between nuclei are fixed to a high approximation: they are subject to temperature dependent fluctuations, but their average positions are essentially unaltered by the electronic signals conveyed to the transistors. It is the electron band structure that is changed via a change in electric potential difference between the base and the conductor, which alters the Fermi surface and depletion zone and hence alters the flow of carriers (electrons and holes).\\

In both cases it is the exquisitely shaped physical context that enables it to happen: here, the precise design and manufacture of the transistors;  in that case, the precise shape of the biomolecules that implement logical choices as explained in \cite{Ellis_Koppel}. In that case it is a result of natural selection and developmental processes; in this case, a result of intelligent design \cite{SimonHA} and precision manufacture: a developmental process occurs whereby physicists and engineers try out all sorts of options to see what works best, leading to very precise doping of semiconductors, precise lithography to create integrated circuits, and so on (\cite{Mellis}:25-29,35-37). The computers and transistors would not exist without this evolutionary process,  
which results 
in the existence of physical products such as digital computers, whose existence cannot be accounted for by physics alone \cite{Ellis2005}. 

\subsection{The outcome: all levels are equally real}\label{sec:levels_all_real}

In the case of biology, all emergent levels are equally real: because of downward causation, they are all causally effective in terms of their own logic, as emphasized by 
Noble \cite{Noble_2012}. Simultaneous same-level causation takes place at every level, with no level being privileged over any other. In the case of digital computers,  each physical level in the implementation hierarchy (Table 2) has real causal powers as expressed in terms of the variables relevant at that level. Causality at the different physical levels is coordinated via a combination of upwards emergence and downwards constraint \cite{Ellis2016}. 
Furthermore, each virtual machine in the logical hierarchy  (\cite{Tannen} and Table 1) is as real as each other; they are all equally causally effective. It is for convenience that we write programs in terms of higher level variables, which are more attuned to the way the human mind thinks. We could in principle write them at any level!\\

    This equal effectiveness of all higher levels is the explicit result of a largely uncelebrated aspect of computers: the development of \href{https://en.wikipedia.org/wiki/Compiler}{compilers} and \href{https://en.wikipedia.org/wiki/Interpreter_(computing)}{interpreters} that translate logic representations downwards between levels \cite{Aho}, without which computers would be largely unusable.\footnote{Try writing a complex program in \href{https://en.wikipedia.org/wiki/Assembly_language}{Assembly language} (\cite{Tannen}:507-521), or much worse, \href{https://en.wikipedia.org/wiki/Machine_code}{ Machine code}! The name of Grace Hopper should be up there  with the panoply of computer greats such as Charles Babbage, Ada Lovelace, Alan Turing, and John von Neumann: see Wikipedia, \href{https://en.wikipedia.org/wiki/Grace_Hopper}{`Grace Hopper'}.} This is the explicit machinery that acts downward to enable abstract high level programs to control flows of electrons in transistors in such a way as to represent abstract logic. In physics terms, the logic of the relevant algorithm is represented by the time dependent pattern of change of the potential energy term (\ref{HV}) in the electron Hamiltonian,  in accord with the digital representation of the algorithm in machine  code. 

\subsection{The  social connection}\label{sec:social_connection}

The high level programs and data are thus represented in machine code, which is realised in physical terms as a time sequence of current flows at the electron level that implements the algorithm logic. But it is 
individual or social needs or purposes that control what programs are actually written and 
change the world, such as Google, Facebook, Paypal, Uber, AirBnB, and so on \cite{Thompson}. Ethical issues necessarily arise, for example as regards how social media such as Facebook use algorithms based in social engineering to increase impact factors \cite{Bissell}, so ethical values are important causal factors in program design. 
The underlying values and understandings of meaning that guide social interactions are in fact the highest level causal factors in computer usage: they are key determinants in deciding what needs will be met in what way by what algorithms. However electron flows make it happen.\\

 The results of this paper have possible implications for arguments regarding emergence and reductionism in the case of the mind/brain (\cite{Hohwyand}, \cite{Humph}), and specifically those arguments against free will based in the alleged causal completeness of physics at the bottom level, related in a key way to the possibility of responsible action. We will not pursue those issues here; we simply comment that the results of \S\ref{sec:causal_power}, combined with those of \cite{Ellis_Koppel}, are potentially relevant to the argument.  
 
\vspace{0.2in}
-------------------------------------------------------------------

\textbf{Acknowledgements} We thank Steven Simon and Oxford University Press for permission to reproduce Figures 2 and 3 from \cite{Simon}. 
 This project was completed while both authors were visiting the Quantum Research Group at the University of KwaZulu Natal (UKZN), and we thank 
 Francesco Petruccione for his hospitality at UKZN and support from his  research grant number 64812: National Research Foundation (South African Research Chair).
\\

We thank an anonymous referee for very helpful comments on a previous version of this paper.

------------------------------------------------------------------

------------------------------------------------------------------------------------------------------------\\

Version: 2019/10/18/Foundations of Physics

\end{document}